\documentclass[%
 reprint,
superscriptaddress,
 amsmath,assume,
 aps,
]{revtex4-2}

\usepackage{graphicx}
\usepackage{dcolumn}
\usepackage{bm}
\usepackage{amsmath}
\usepackage{physics}
\usepackage{mathtools}
\usepackage{multirow}
\usepackage{cancel}
\usepackage{amssymb}
\usepackage[super]{nth}
\usepackage{graphicx}
\usepackage[hidelinks,colorlinks=true,linkcolor=blue,urlcolor=cyan,citecolor=green]{hyperref}
\usepackage{color}
\usepackage{qcircuit}
\usepackage{caption}
\usepackage{subcaption}
\usepackage[utf8]{inputenc}
\usepackage{natbib}
\usepackage{verbatim}

\newcommand{\mbf}{\mathbf}
\newcommand{\mbb}{\mathbb}
\newcommand{\mc}{\mathcal}

\renewcommand{\op}[2]{|#1\rangle\langle #2|}

\definecolor{cool_green}{rgb}{0.0, 0.5, 0.0}

\begin{document}

\preprint{APS/123-QED}

\title{Implementing Quantum Secret Sharing on Current Hardware }%

\author{Jay Graves}
 \email{jay.graves@yale.edu}
\affiliation{Department of Physics, Morehouse College, Atlanta, GA 30314}%
\affiliation{Coordinated Science Laboratory\\ Department of Electrical and Computer Engineering, University of Illinois at Urbana-Champaign, Urbana, IL 61801
}%
\affiliation{Department of Physics, Yale University, New Haven, CT 06511}%
\author{Mike Nelson}%
 \email{minelson@illinois.edu}
\affiliation{Coordinated Science Laboratory\\ Department of Electrical and Computer Engineering, University of Illinois at Urbana-Champaign, Urbana, IL 61801
}%
\author{Eric Chitambar}%
 \email{echitamb@illinois.edu}
\affiliation{Coordinated Science Laboratory\\ Department of Electrical and Computer Engineering, University of Illinois at Urbana-Champaign, Urbana, IL 61801
}%



\begin{abstract}
Quantum secret sharing is a cryptographic scheme that enables the secure storage and reconstruction of quantum information. While the theory of secret sharing is mature in its development, relatively few studies have explored the performance of quantum secret sharing on actual devices. In this work, we provide a pedagogical description of encoding and decoding circuits for different secret sharing codes, and we test their performance on IBM's 127-qubit \textit{Brisbane} 
 system. We evaluate the quality of the implementation by performing a SWAP test between the decoded state and the ideal one, as well as by estimating how well the code preserves entanglement with a reference system. The results indicate that a ((3,5)) threshold secret sharing scheme and a non-threshold 7-qubit scheme perform similarly based on the SWAP test and entanglement fidelity, with both attaining a roughly \mbox{70--75\%} pass rate on the SWAP test for the reconstructed secret. We also investigate one implementation of a ((2,3)) qutrit threshold scheme and find that it performs the worst of all, which is expected due to the additional number of multi-qubit gate operations needed to encode and decode qutrits. A comparison is also made between schemes using mid-circuit measurement versus delayed-circuit measurement.

\end{abstract}

\maketitle


\section{\label{sec:level1}Introduction}

Securely storing and distributing information is one of the oldest and most important communication tasks. With~the advent of quantum computing, traditional cryptographic techniques are being challenged in new and disruptive ways. At~the same time, quantum computing also offers a pathway for realizing new cryptographic schemes with potentially stronger security guarantees than their classical counterparts. One such example is quantum secret sharing (\textit{QSS}) \cite{Cleve-1999a}, which offers a solution to the problem of distributing secrets among multiple~parties.

In classical cryptography, Blakley~\cite{Blakley-1979a} and Shamir~\cite{Shamir-1979a} were the first to formally introduce the theory of secret sharing. The~most well-known protocols are ($k$, $n$) threshold schemes, in~which a secret is divided into $n$ shares; any $k$ of those shares can be used to reconstruct the secret, but~any fewer number contains no information about the secret. More generally, one constructs codes in which just certain subsets of shares can recover the secret, as~specified by the code's access structure. Hillery \textit{et al.} later observed that quantum states could be used for building threshold schemes by encoding a classical secret into a multipartite entangled state~\cite{Hillery-1999}. Unlike traditional secret sharing protocols that rely on computational hardness assumptions, the~security of ideal quantum-based schemes depends only on the physical principles of quantum mechanics. Going one step further, Cleve \textit{et al.} introduced fully quantum secret sharing protocols~\cite{Cleve-1999a}, which involve the distribution and reconstruction of a quantum secret, i.e.,~the superposition state of some quantum system. Since these initial works, the~theory of QSS has received heavy development and a wide range of protocols have been proposed~\cite{Gottesman-2000a, smith-2000, Nascimento-2001a, imai2003, Markham-2008a, Sarvepalli-2011a, Gheorghiu-2012a, Fortescue-2012a, Yang-2013a, Gheorghiu-2013a, Marin-2013a, Li-2013a, Ouyang-2017a, Lipinska-2020a, Senthoor-2022a, Ouyang-2023a}. Complementing this theory work, there has been experimental progress on implementing different QSS schemes, primarily in photonic systems~\cite{Tittel-2001a, Gaertner-2007a, Hao-2011a, Wei-2013a, Bell-2014a, Lu-2016a, Williams-2019a, Lee-2020a, deOliveira-2020a}.

These experimental demonstrations involve setups that are designed specifically for the task of secret sharing. However, any functioning quantum computer should be able to run basic QSS protocols. In~fact, since every QSS scheme utilizes a type of error-correcting code~\cite{Cleve-1999a}, secret sharing is a natural functionality to the demand of a quantum computer on the road toward fault-tolerance. Furthermore, while secret sharing is often described in the scenario of multiple parties separated by some large spatial distance, it also has applications in much smaller computational settings. For~example, one can imagine a modular quantum computing architecture in which sensitive data gets distributed onto different zones to minimize a potential data breach. Motivated by these considerations, we investigate the performance of quantum secret sharing codes on current quantum computing hardware. As~the simplest possible benchmark, we consider a single-qubit secret, checking whether privacy is attained after encoding and whether the secret is successfully recovered after~decoding.

In the following, we begin by reviewing the basic elements of QSS and describing the three codes that we implement; a more comprehensive overview can be found in the work of Gottesman~\cite{Gottesman-2000a}. In~Section~\ref{Sect:Preliminaries}, we outline our encoding and decoding methods in more detail and also explain our figures of merit for the code performance. The~codes were implemented on IBM's 127-qubit Brisbane superconducting processor, and~the results are presented in Section~\ref{Sect:results}. In~particular, we find that the ((3,5)) QSS scheme and the 7-qubit scheme perform very similarly in terms of passing the SWAP test with pass rates ranging between 65 and 80\% in both the simulations and real experiments. The simulations predict that the ((3,5)) QSS scheme should preserve entanglement better; however, it remains unclear using actual hardware. We also find that the use of higher optimization levels within Qiskit enhances the performance of the QSS protocols and that classical feed-forward is more efficient than fully coherent decoding for QSS schemes. We firmly expect that the performance can be improved even further by employing other error mitigation techniques, such as dynamical decoupling~\cite{Suter-2016a}. 

Those familiar with QSS can skip immediately to Section~\ref{Sect:results} for a summary of our results. A~pedagogical introduction to QSS is provided in Sections~\ref{Sect:Preliminaries} and \ref{Sect:Examples}.  Our hope is that this not only makes this paper more accessible but also useful as a basic reference for constructing QSS~circuits.

\section{Preliminaries}

\label{Sect:Preliminaries}

Suppose that $\ket{\psi}=\alpha\ket{0}+\beta\ket{1}$ is an arbitrary qubit state that some ``dealer'' wishes to distribute to $n$ parties. If~each party itself is given a qubit system, then the dealer performs an encoding isometry $V:\mbb{C}^2\to\mbb{C}^n$, which maps $\ket{\psi}$ into the $n$-qubit state
\[\ket{\psi}\mapsto V\ket{\psi}=\ket{\psi}_L=\alpha\ket{0}_L+\beta\ket{1}_L.\]
Here, $\ket{i}_L\in\mbb{C}^n$ (for $i\in\{0,1\}$) are states forming a basis for a logical qubit in $\mbb{C}^n$. Letting $2^{[n]}$ denote the power set of $[n]:=\{1,\cdots,n\}$, every QSS scheme is defined by an access structure $\Gamma\subset 2^{[n]}$ such~that 
\begin{enumerate}
    \item[(i)] Each $S\in 2^{[n]}\setminus \Gamma$ is called unauthorized and satisfy
\begin{equation}
\omega^{S}=\tr_{\overline{S}}(\op{\psi}{\psi}_L),
    \end{equation} 
    where $\overline{S}$ denotes the set complement of $S$ and $\omega^S$ is some fixed state for systems $S$ that is independent of $\ket{\psi}$, i.e.,~no statistical information about the relative values of $|\alpha|$ and $|\beta|$ can be obtained.
    \item[(ii)] Each $S\in \Gamma$ is called authorized, and~there exists a decoder $\mc{D}_S$ such that
\begin{equation}
    \label{Eq:decoder}
        \op{\psi}{\psi}=\mc{D}_S(\tr_{\overline{S}}(\op{\psi}{\psi}_L)).
    \end{equation}
\end{enumerate}
A $((k,n))$ threshold scheme is a special type of QSS in which $\Gamma$ consists of all subsets having $k$ or more parties.
The first two QSS schemes we investigate are built from [[$n,m,d$]] 
 qubit stabilizer codes. Such codes can correct $t=\lfloor(d-1)/2\rfloor$ general errors and $d-1$ erasure errors~\cite{9781107002173}. We specifically consider the [[5,1,3]] (``five-qubit'') code~\cite{Laflamme-1996a}  and the [[7,1,3]] (``Steane'') CSS code~\cite{Calderbank-1996a, Steane-1996a}, both of which correct two erasure errors on any subset of qubits. The~[[5,1,3]] code admits a ((3,5)) threshold scheme, and~it is a maximum distance separable code, i.e.,~it satisfies the quantum Singleton bound, $n-m \geq 2(d-1)$, with~equality, while the Steane code does~not.

While the Steane code is a $((5,7))$ threshold scheme, it can also correct three or even four erasure errors located on certain subsets of qubits.  In~this paper, we explore the more intricate QSS access structure that arises from the Steane code. Finally, we test the exemplifying ((2,3)) qutrit QSS scheme presented in Ref.~\cite{Cleve-1999a}, which is a [[3,1,2]]$_3$ CSS code that can correct a single erasure error and detect a single general error (but cannot correct the latter~\cite{majumdar-2020}).

\subsection{General Encoding}

The encoding procedure for stabilizer codes can be implemented in different ways. One method is to consider the evolution of the stabilizer. If~$\{g_i\}_{i=1}^{n-m}$ is an independent set of generators for an [[$n,m,d$]] stabilizer code and $\{\overline{Z}_i\}_{i=n-m+1}^{n}$ are logical $Z$ operators, then there always exists a Clifford unitary $U$ that maps $\{Z_i\}_{i=1}^n$, stabilizers of the initial $n$-qubit state $\ket{0}^{\otimes n}$, to~$\{g_i\}_{i=1}^{n-m}\cup \{\overline{Z}_i\}_{i=n-m+1}^{n}$, stabilizers of the logical state $\ket{0}_{L}$; i.e.,~$UZ_iU^\dagger =g_i $ for $i=1,\cdots n-m,$ and $UZ_iU^\dagger =\overline{Z}_i$, for~$i=n-m+1,\cdots, n$. One then needs to find an implementation of $U$ that is compatible with the native gate set for whatever quantum computing hardware device is being used. Our results were obtained on the IBM Brisbane machine, whose standard gate set consists of qubit rotations about the $\hat{z}$-axis ($R_Z(\phi)$), $\pi$ and $\pi/2$ rotations about the $\hat{x}$-axis ($X$ and $\sqrt{X}$), and~the two-qubit echoed cross-resonance gate (ECR) gate. The~ECR gate is equivalent to a CNOT up to single-qubit pre-rotations. The~qubit rotation matrices about each of the standard axes on the Bloch sphere are \mbox{given by}
\begin{align}
R_X(\theta) &= \begin{bmatrix}
\text{cos}(\theta/2) & -i\text{sin}(\theta/2) \\
-i\text{sin}(\theta/2) & \text{cos}(\theta/2) 
\end{bmatrix}\notag\\
R_Y(\theta) &= \begin{bmatrix}
\text{cos}(\theta/2) & -\text{sin}(\theta/2) \\
\text{sin}(\theta/2) & \text{cos}(\theta/2) 
\end{bmatrix}\notag\\
R_Z(\phi) &= \begin{bmatrix}
e^{-i\phi/2} & 0 \\
0 & e^{i\phi/2}
\end{bmatrix},\notag
\end{align} 
and we can realize the $R_X(\theta)$ and $R_Y(\theta)$ gates using IBM Brisbane's native gate set using the following relations:
\begin{align}
R_X(\theta) &= R_Z\left(\frac{\pi}{2} \right) \sqrt{X} R_Z(\theta + \pi) \sqrt{X} R_Z\left(\frac{\pi}{2} \right), \notag\\
R_Y(\theta) &= R_Z(\pi) \sqrt{X} R_Z(\theta + \pi) \sqrt{X}. \notag
\end{align} 

The design of all circuits was accomplished using Qiskit. Since Qiskit only supports qubit encodings, to~implement the $((2,3))$ qutrit code, we essentially embedded each qutrit into a two-qubit system, as~explained in Section~\ref{Sect:encoding_qutrit}.

\subsection{General Decoding}

\begin{figure*}
  \includegraphics[width=.8\textwidth,height=3.5cm]{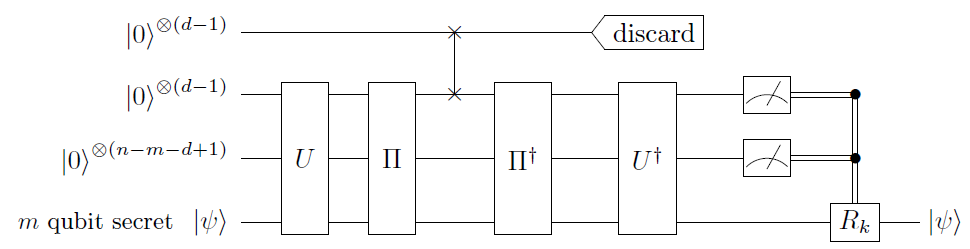}
  \caption{A general threshold QSS circuit implementation using an [[$n,m,d$]] stabilizer code. An~arbitrary $m$-qubit secret $\ket{\psi}$ is encoded into $n$ qubits by a unitary $U$. A~permutation $\Pi$ is performed to select an unauthorized set (erased set $E$) of $d-1$ qubits that get discarded. A~fresh set of qubits is swapped in place of the latter qubit, and~the encoding map is reversed. Finally, an~error syndrome is obtained by measuring the $n-m$ qubits in the computational basis, and the appropriate correction $R_k$ is performed on the unmeasured qubits to recover the secret $\ket{\psi}$. Note that the error correction $R_k$ will depend on the permutation $\Pi$.}
  \label{Fig:encode-decode-circuit}
\end{figure*}

The general encoding procedure just described takes an initial $m$-qubit secret $\ket{\psi}$ prepared in registers $n-m+1,\cdots, n$ and maps it into an $n$-qubit logical state. In~QSS, the~secret is recovered after discarding an unauthorized subset of parties. To~explain how this is accomplished in practice, let us first describe in more detail the error correction procedure for an erasure event. Suppose it is known that some specific physical qubits are lost in a computation or protocol. Some physical reasons for loss are described in Ref.~\cite{Grassl-1997a}, but~in QSS, the loss is an artifact of the task itself: an authorized subset of parties participate collaboratively to recover the secret, and~the remaining qubits are effectively lost since they are held by non-participating parties. To~recover the secret, the~authorized parties replace the lost qubits with fresh qubits, each in some initial state $\ket{0}$. When qubit $Q_i$ is lost and replaced by qubit $A_i$, its mathematical description is described by the completely positive trace-preserving (CPTP) map
\begin{align}
    \mc{D}^{Q_i\to A_i}(X^{Q_i})&=\tr[X^{Q_i}]\op{0}{0}^{A_i}\notag\\
    &=\op{0}{0}X^{Q_i}\op{0}{0}^{A_i}+\op{0}{1}X^{Q_i}\op{1}{0}^{A_i},\notag
\end{align}
where we have provided one representation of the map in terms of Kraus operators $K_0={}^{A_i}\!\op{0}{0}^{Q_i}$ and $K_1={}^{A_i}\!\op{0}{1}^{Q_i}$. If~a code can correct arbitrary errors on qubit $Q_i$, then it can also correct the erasure map $\mc{E}^{Q_i\to A_i}$ by transferring the error correction procedure from system $Q_i$ to $A_i$. The~overall effect is that qubits $Q_i$ and $A_i$ get swapped in between the encoding and decoding procedure (see Figure~\ref{Fig:encode-decode-circuit}).  

The standard method for decoding stabilizer codes is to first measure a complete set of generators for the stabilizer, the~outcomes of which uniquely identify an error syndrome. The~error is then reversed, and the inverse of the encoding gate, $U^\dagger$, is applied to recover the secret $\ket{\psi}$. More explicitly, if~$\{g_i\}_i$ are generators for the stabilizer and $\{E_k\}_k$ are correctable Pauli errors, then each $E_k$ has the error syndrome determined by the bit string 
 $\mbf{b}_k=(b_{k,i})_i$, such that $E_k g_i E^\dagger_k=(-1)^{b_{k,i}}g_i$. The~error syndrome is obtained by measuring each of the generators $g_i$, and~if the syndrome is $\mbf{b}_k$, then the error can be corrected by performing $E_k^\dagger$ or any other $E_l^\dagger$ with the same syndrome. Afterward, $U^\dagger$ is applied to recover the secret $\ket{\psi}$ in registers $n-m+1,\cdots, n$. Equivalently, since
\begin{align}
(-1)^{b_{k,i}}U^\dagger g_i U&=U^\dagger  E_k U(U^\dagger g_i U)U^\dagger E_k^\dagger U\notag\\
\Leftrightarrow\;\;\quad(-1)^{b_{k,i}}Z_i&=(U^\dagger  E_k U )Z_i (U^\dagger E_k^\dagger U),
\end{align}
one can first evolve the circuit by $U^\dagger$ prior to the syndrome measurement. The~syndrome $\mbf{b}_k$ can then be determined by measuring each of the qubits $l=1,\cdots,n-m$ in the computational basis, and~the corresponding error correction is facilitated by performing $U^\dagger E_k^{\dagger} U$. For~a distance $d$ stabilizer code and arbitrary subset $T$ of $d-1$ parties, it suffices to consider just the set of Pauli errors $\{E_k\}_k=\{\mbb{I}, X,Y,Z\}^{\times (d-1)}$ acting on qubits in $T$.  Notice that since $U$ is a Clifford gate, each $U^\dagger E_k^{\dagger} U$ will also be a local Pauli operator. We let $R_k$ denote the part of $U^\dagger E_k^\dagger U$ acting on the $m$-qubit system recovering the secret, and~for simplicity, we ignore resetting the other $n-m$ systems to their original state $\ket{0}^{\otimes (n-m)}$. The~overall encoding and decoding scheme is depicted in Figure~\ref{Fig:encode-decode-circuit}.

\subsection{Initializing a secret}

A $d$-dimensional quantum secret is an arbitrary state in $\mbb{C}^d$. 
For qubits we can parametrize the secret as $\ket{\psi}$ = $\alpha\ket{0} + \beta\ket{1} = \text{cos}(\theta/2)\ket{0} + \text{sin}(\theta/2)e^{i\phi}\ket{1}$. The~use of rotation gates, specifically $R_X(\theta)$ and $R_Z(\phi)$, rotates the state around the Bloch sphere depending on the angles $\theta$ and $\phi$. This allows us to initialize a random secret when we randomize the input angles of the rotation gates, as~depicted in Figure~\ref{fig:qubit_sec}. 

\begin{figure}[b]
    \begin{subfigure}{0.45\textwidth}
        \caption{}
        \[
        \qquad \qquad \qquad \Qcircuit @C=1.5em @R=1em  {
        \lstick{\ket{\psi}\;\;\equiv\;\; \ket{0}}& \gate{R_X(\theta)}&\gate{R_Z(\phi)}&\qw  \\
        }
        \]
        \label{fig:qubit_sec} 
    \end{subfigure}
    \hfill
    \begin{subfigure}{0.45\textwidth}
        \caption{}
        \[
        \qquad \qquad \Qcircuit @C=1em @R=1em {
        \lstick{\ket{0}}&\gate{R_Y(\theta_1)} & \targ & \gate{R_Y(\theta_1)} & \targ & \qw \\
        \lstick{\ket{0}}&\gate{R_Y(\theta_2)} & \ctrl{-1} & \qw & \ctrl{-1} & \qw
        \inputgrouph{1}{2}{1.3em}{\ket{\psi}\;\;\equiv \qquad}{3.2em}
        }
        \]
        \label{fig:qutrit_sec}
    \end{subfigure}
\caption[Two numerical solutions]{(a) Initialization of arbitrary qubit quantum state. (b) Initialization of arbitrary qutrit quantum state using qubits.}
\end{figure}

More specifically, we arranged $\theta$ and $\phi$ into arrays with intervals $[0, \pi]$ and $[0, 2\pi]$, respectively, both with a step of $1^{\circ}$. With~each circuit job, a~random angle is selected from each array and used for the angles of the rotation gates, thus initializing an arbitrary quantum state or~secret.

For the qutrit scheme, we embed a three-level system into the space of two qubits, $\ket{\psi} \in \mbb{C}^3 \rightarrow
\ket{\phi} \in \mbb{C}^2 \otimes \mbb{C}^2$, while ignoring one of the basis states of $ \mbb{C}^2 \otimes \mbb{C}^2$. Specifically, we map a qutrit into two qubits using the mapping
\begin{align}
 \ket{0}\leftrightarrow \ket{00},\quad\ket{1}\leftrightarrow\ket{01},\quad\ket{2}\leftrightarrow\ket{10}.   \label{Eq:qutrit-encoding}
\end{align}
 
\noindent Under the qutrit embedding of Equation \eqref{Eq:qutrit-encoding}, we can initialize an arbitrary real qutrit state $\ket{\psi}$ using $R_Y(\theta)$ and CNOT gates as depicted in Figure~\ref{fig:qutrit_sec}. It is straightforward to verify that $\ket{\psi} = \alpha\ket{0} + \beta\ket{1} + \gamma\ket{2} \coloneqq \alpha\ket{00} + \beta\ket{01} + \gamma\ket{10}$, where $\alpha=\cos(\theta_2/2)$, $\beta=\sin(\theta_2/2)\cos(\theta_1)$, and~$\gamma=\sin(\theta_2/2)\sin(\theta_1)$.


\subsection{Measures of performance}

\label{Sect:Performanc_Measures}

We consider two different ways of evaluating how well a quantum circuit implements a QSS scheme. The~first uses the SWAP test, which is a computational primitive shown in Figure~\ref{fig:graph2}. The~probability of measuring $\ket{0}$ is given by $1/2(1+|\ip{\psi}{\phi}|^2$), which becomes unity if the states are the same. Hence, our test will involve (i) randomly selecting angles $\theta$ and $\phi$ for the initial qubit state $\ket{\psi}$, (ii) preparing two copies of $\ket{\psi}$, (iii) running the QSS encoding and decoding on one of the copies, and~(iv) performing the SWAP test with the original unencoded state.

\begin{figure}[b]
\[
\qquad \Qcircuit @C=1.5em @R=1.5em  {
\lstick{\ket{0}} & \gate{H} & \ctrl{2} & \gate{H} & \meter & \cw \\
\lstick{\ket{\phi}} & \qw & \qswap & \qw & \qw & \qw \\
\lstick{\ket{\psi}} & \qw & \qswap & \qw & \qw & \qw
}
\]
\caption{\label{fig:graph2} Quantum SWAP test circuit diagram.} 
\end{figure} 

The second quantifier of performance for a QSS involves the entanglement fidelity, which for a general quantum channel $\mc{N}$ is defined as
\begin{equation}
F_e(\mc{N}) \coloneqq \langle \Phi^+|(\text{id} \otimes \mc{N})(\op{\Phi^+}{\Phi^+}) \ket{\Phi^+},
\label{eq:entfidel}
\end{equation}
where $\ket{\Phi^+}=\sqrt{1/2}(\ket{00}+\ket{11})$. Intuitively, the~entanglement fidelity measures how well the channel $\mc{N}$ preserves the maximally entangled state $\ket{\Phi^+}$ when acting on one of the subsystems. In~our case, we consider $\mc{N}$ as the concatenation of the QSS encoding $V$ and the decoding $\mc{D}_S$ after erasing shares in subset $S$. Ideally, $\ket{\Phi^+}$ should be perfectly preserved after encoding and decoding. For~$d$-dimensional codes, we replace $\ket{\Phi^+}$ with $\ket{\Phi^+_d}=\tfrac{1}{\sqrt{d}}\sum_{i=0}^{d-1}\ket{ii}$ in Equation \eqref{eq:entfidel}.

To estimate the entanglement fidelity for a given subset erasure, we perform quantum state tomography on the output state when starting with $\ket{\Phi^+}$. Quantum state tomography (QST) is a method for experimentally reconstructing the quantum state from measurement data. QST requires multiple measurements of the quantum system in different bases. For~a system of $n$ qubits, there are $3^n$ possible measurement bases (corresponding to measurements along the $X$, $Y$, and~$Z$ axes for each qubit). Qiskit offers a native QST command, called StateTomography, which we employed in collecting our~data.

\subsection{Error mitigation}

The digital units of a quantum computer (qubits, typically) are very fragile and subject to noise and errors in computation. In~the long-term vision of quantum computers, fully fault-tolerant devices can suppress noise and errors to an arbitrarily small degree. However, such low-error-rate devices require substantial physical resources. In~the present regime of so-called Noisy Intermediate Scale Quantum (NISQ) computation, various strategies for error mitigation have been suggested and demonstrated. These approaches provide some benefits beyond fault-tolerance, without~scaling up physical device~requirements. 

One such approach is matrix-free measurement mitigation (Mthree or M3), which specifically focuses on measurement errors~\cite{PRXQuantum.2.040326}. M3 relies on probabilistic methods to estimate and correct errors. These methods can introduce approximations that may not fully capture the complexity of the noise profile, leading to less accurate corrections compared to full matrix methods. Moreover, the~calibration phase in M3 typically involves measuring only a subset of possible error configurations (e.g., single-qubit or pairwise errors). This can miss higher-order correlations between errors on multiple qubits, reducing the effectiveness of the mitigation. Otherwise, M3 is an efficient routine and is a promising technique for obtaining accurate experimental results for large quantum systems, where the number of possible measurement outcomes grows exponentially. A~more detailed description of M3 can be found in Ref.~\cite{PRXQuantum.2.040326}.

\section{Specific Examples}
\label{Sect:Examples}

\subsection{((3,5)) QSS using the five-qubit code}
\label{encoding_5}

\begin{figure}[t]
  \includegraphics[scale=0.29]{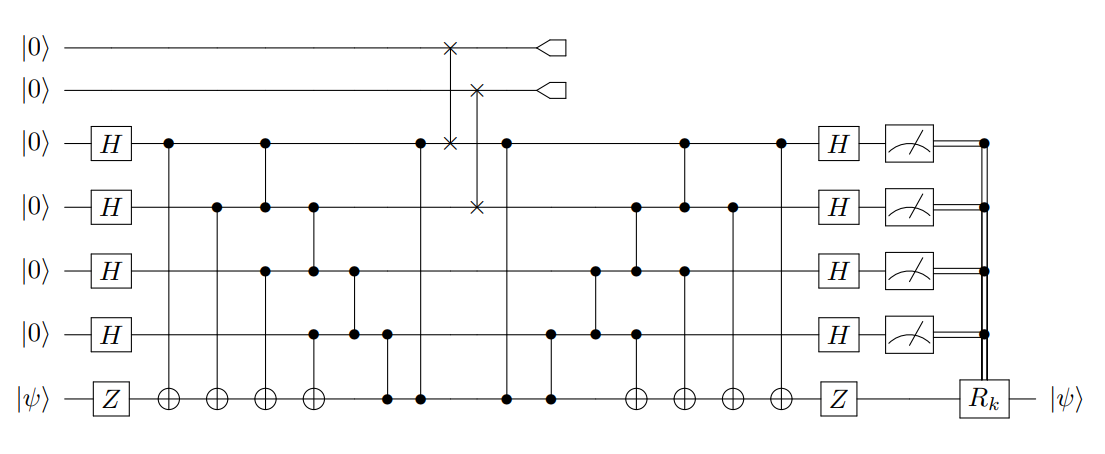}
  \caption{A circuit implementation of a five-qubit QSS protocol.  }
  \label{Fig:5-qubit}
\end{figure}

We now apply the general method outlined in Section~\ref{Sect:Preliminaries} to the [[$5,1,3$]] five-qubit code. Generators for the stabilizer of this code are given in Table~\ref{table:table3}. Our first task is to identify the encoding unitary $U$ that maps the initial stabilizers $\{Z_i\}_{i=1}^5$ to the $\{G_i\}_{i=1}^4\cup\{\overline{Z}\}$ in Table~\ref{table:table3}. This mapping can be constructed by observing that the generators of the five-qubit code closely resemble the stabilizers for the five-qubit ring state~\cite{Markham-2008a}. The~latter is represented by a ring graph, and~it is generated by performing a controlled-$Z$ (CZ) between adjacent nodes on the graph, each initially in the state $\ket{+}$ (and hence stabilized by $X$). Sequentially applying (i) the same CZ gates on the five-qubit code, (ii) CNOT on the appropriate pairs of qubits, and~(iii) Hadamard gates to qubits 1--4 transforms the generators as
\begin{align}
\label{Eq:stabilizer-1}
    \begin{pmatrix}G_1\\G_2\\ G_3\\ G_4\\\overline{Z}\end{pmatrix}\mapsto\begin{pmatrix}XIIXI\\IXIIX\\XIXII\\IXIXI\\IIIIZ\end{pmatrix}\mapsto\begin{pmatrix}XIIII\\IXIII\\IIXII\\IIIXI\\IIIIZ\end{pmatrix}\mapsto\begin{pmatrix}ZIIII\\IZIII\\IIZII\\IIIZI\\IIIIZ\end{pmatrix}.
\end{align}
Here, we have used the fact that CNOT converts $X_cX_t\mapsto X_c$ and $Z_cZ_t\mapsto Z_t$ for the control ($c$) qubit and target ($t$) qubit.  Two equivalent circuits realizing the CNOT transformations are the following:
\[
\Qcircuit @C=.5em @R=.6em  {
& \targ&\ctrl{3}&\qw&\qw&\qw\\
 &\qw&\qw&\targ&\ctrl{3}&\qw\\
&\ctrl{-2}&\qw&\qw&\qw&\qw\\
&\qw&\targ&\ctrl{-2}&\qw&\qw\\
&\qw&\qw  &\qw&\targ&\qw\\
} \quad\equiv \quad
\Qcircuit @C=.5em @R=1em  {
& \ctrl{4}&\qw&\qw&\qw&\qw\\
&\qw&\ctrl{3}&\qw&\qw&\qw\\
&\qw&\qw&\ctrl{2}&\qw&\qw\\
&\qw&\qw&\qw&\ctrl{1}&\qw\\
&\targ&\targ&\targ&\targ&\qw\\
}.
\]
\noindent Performing Equation \eqref{Eq:stabilizer-1} in reverse defines the encoding unitary $U$, as~depicted in \mbox{Figure~\ref{Fig:5-qubit}}. Here, we have chosen the permutation $\Pi$ among the unauthorized set to be trivial \mbox{for~convenience.}

The next step is to identify the Pauli correction $R_k$ for each error syndrome $\mbf{b}_k$. This is performed by observing how the Pauli errors on the first and second qubits evolve during the application of $U^\dagger$. The~necessary correction $R_k$ on the fifth qubit is summarized in Table~\ref{Tab:correction-5-qubit}. Finally, we perform the fidelity tests to see how well the procedure works at recovering an arbitrary initial secret $\ket{\psi}$. Our results are presented in Section~\ref{Sect:results}.

\subsection{QSS using the Steane code}
\label{encoding-steane-57}

We now apply the same general method to the [[7,1,3]] Steane code. Generators for the stabilizer of this code are given in Table~\ref{table:table3}. As~stated previously, CNOT converts $X_cX_t\mapsto X_c$ and CZ converts $Z_cZ_t\mapsto Z_t$. Therefore, by~applying CNOT to the appropriate pairs of qubits, followed by Hadamard gates to qubits $5,6,7$, and a multiplication of qubit $1$ by $Z$, yields the sequence of transformations
\begin{align}
\label{Eq:stabilizer-4}
    \begin{pmatrix}G_1\\ G_2\\ G_3\\ G_4\\ G_5\\ G_6\\\overline{Z}\end{pmatrix}\xmapsto{}\begin{pmatrix}
    IIIIXXX\\
    IIIIIXX\\
    IIIIXIX\\
    IIIZIII\\
    IZZIIII\\
    IIZIIII\\
    IZZZIII
    \end{pmatrix}\xmapsto{}
        \begin{pmatrix}
    ZIIIZZZ\\
    ZIIIIZZ\\
    ZIIIZIZ\\
    ZIIZIII\\
    ZZZIIII\\
    ZIZIIII\\
    ZZZZIII
    \end{pmatrix}.
\end{align}

\noindent By adding together different combinations of the generators, it is straightforward to verify that this stabilizes the state $\ket{0}^{\otimes 7}$, as~desired.

\begin{figure}[h]
        \includegraphics[scale=0.33]{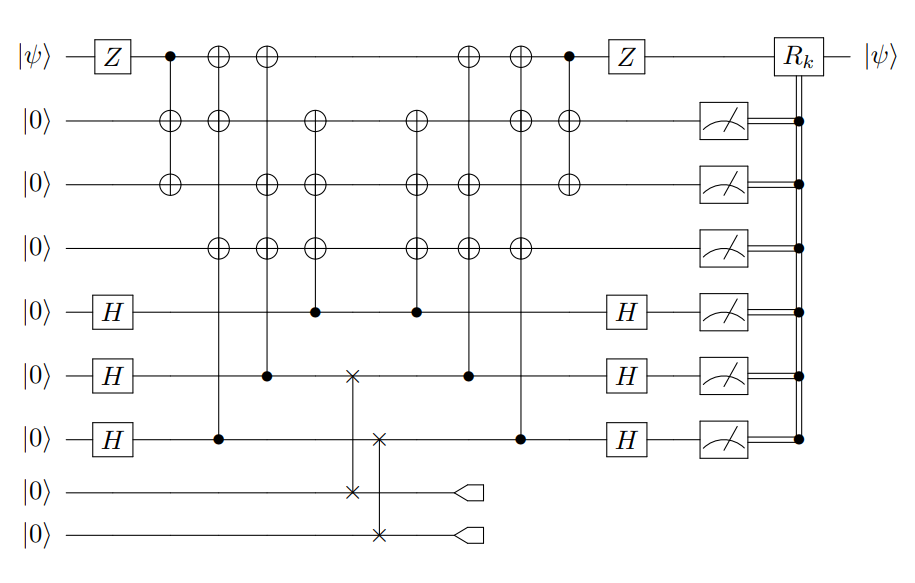}
\caption[Two numerical solutions]{A circuit implementation of a Seven-qubit (Steane) QSS protocol with two qubits being erased, $|E|=2$.}
\label{Fig:7-qubit} 
\end{figure}

As was the case with the ((3,5)) scheme, applying these transformations in reverse defines the encoding unitary, $U$, which maps the initial stabilizers $\{Z_i\}_{i=1}^7$ to the $\{G_i\}_{i=1}^6\cup\{\overline{Z}\}$, as depicted in Figure~\ref{Fig:7-qubit}. We have chosen the unauthorized set to be the sixth and seventh qubits. The~necessary correction $R_k$ on the first qubit is summarized in Table~\ref{Tab:correction-7-qubit}.

\begin{table}[b]
\caption{Stabilizer generators for the 5-qubit (left) and Steane (right) codes.} 
\centering 
\begin{ruledtabular}
\begin{tabular}{c c c c} 
Element & Operator & Element & Operator \\  
\hline 
$G_1$ & $XZZXI$ & $G_1$ & $IIIXXXX$ \\
$G_2$ & $IXZZX$ & $G_2$ & $IXXIIXX$ \\
$G_3$ & $XIXZZ$ & $G_3$ & $XIXIXIX$ \\
$G_4$ & $ZXIXZ$ & $G_4$ & $IIIZZZZ$ \\
$XXXXX$ & $\bar{X}$ & $G_5$ & $IZZIIZZ$ \\
$ZZZZZ$ & $\bar{Z}$ & $G_6$ & $ZIZIZIZ$ \\
& & $\bar{X}$ & $XXXXXXX$\\
& & $\bar{Z}$ & $ZZZZZZZ$\\
\end{tabular}
\end{ruledtabular}
\label{table:table3} 
\end{table}

The Steane code enables a more intricate QSS access structure than just a $((5,7))$ threshold scheme~\cite{Pradeep-2012a}. As~shown in Appendix \ref{App:Steane-access}, the~code can still recover when certain subsets of three or even four qubits are erased (but not all subsets of those sizes). For~CSS codes, where $C^\perp \subseteq C \subseteq \mathbb{F}_q^n$, the~minimal access structure, $\Gamma_m$, is given by the support of the minimal codewords in $C\setminus C^{\perp}$ \cite{Sarvepalli_2009}.
\begin{align*}
    \Gamma^{\rm Steane}_m = \textbf{\{} \{1,2,3\},\{1,4,5\}, \{1,6,7\}, \{2,4,6\}, \\ \{2,5,7\}, \{3,4,7\}, \{3,5,6\} \textbf{\}}
\end{align*}
The encoding remains the same, but~a different recovery is needed. Figure~\ref{Fig:4-7-qubit-circuit} depicts one such circuit in which qubits one through four form an authorized set. The~necessary correction $R_{k}$ on the first qubit for all possible syndromes is presented in Appendix \ref{App:Steane-access}.

\subsection{((2,3)) QSS qutrit code}

\label{Sect:encoding_qutrit}

\begin{figure}[!]
  \includegraphics[scale=0.22]{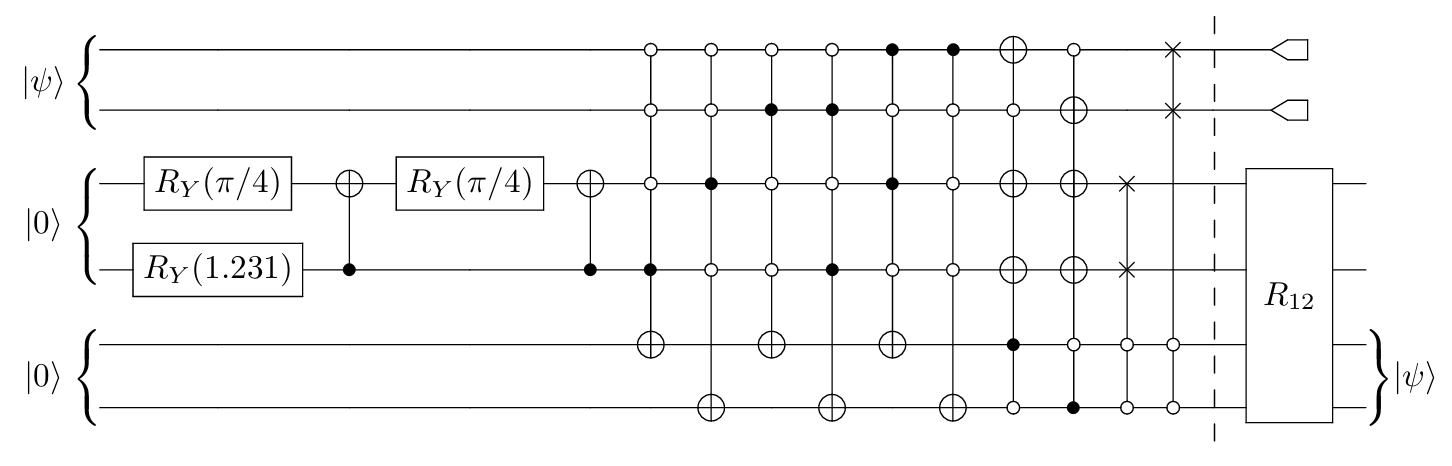}
  \caption{A circuit implementation of a 3-qutrit QSS protocol (using qubits). The~first qutrit is represented by the lower pair of qubits, the~second by the middle pair, and~so~on. }
  \label{Fig:3-qutrit}
\end{figure}

The qutrit scheme involves secrets of the form $\ket{\psi}=\alpha\ket{0} + \beta\ket{1} + \gamma\ket{2}$. The~secret sharing process was laid out by Cleve \textit{et al} \cite{Cleve-1999a}. While qutrit protocols are not ideally suited for qubit-based hardware, it is still possible to implement them. 
In this work, we have adopted the following logical encoding:
\begin{eqnarray}
\label{Eq:qutrit-logical}
\tilde{\ket{\psi}}_L= && \quad \alpha\tilde{\ket{0}}_L + \beta\tilde{\ket{1}}_L + \gamma\tilde{\ket{2}}_L \\
= && \quad \frac{\alpha}{\sqrt{3}} \qty(\ket{000} + \ket{111} + \ket{222}) \nonumber\\
+ && \quad \frac{\beta}{\sqrt{3}} \qty(\ket{012} + \ket{120} + \ket{201}) \nonumber\\
+ && \quad \frac{\gamma}{\sqrt{3}} \qty(\ket{021} + \ket{102} + \ket{210}) \nonumber\\
\nonumber
\end{eqnarray}
The secret can be reconstructed from any two of the three shares. Given the first two shares (for instance), the~recovery unitary $R_{12}$ involves adding the value of the first share to the second (modulo three), and~then adding the value of the second share to the first. This yields the state~\cite{Cleve-1999a}
\begin{eqnarray}
\notag
(R_{12} \otimes I_3)\tilde{\ket{\psi}}_L= \quad \frac{\alpha}{\sqrt{3}} \qty(\ket{000} + \ket{021} + \ket{012}) \\
+  \quad \frac{\beta}{\sqrt{3}} \qty(\ket{112} + \ket{100} + \ket{121}) \nonumber\\
+  \quad \frac{\gamma}{\sqrt{3}} \qty(\ket{221} + \ket{212} + \ket{200}) \nonumber\\
=  \quad \ket{\psi}\otimes \frac{1}{\sqrt{3}}\left( \ket{00} + \ket{12} + \ket{21} \right), \notag
\end{eqnarray}
which recovers the secret in the first qutrit share. The~reconstruction procedure for the other cases is similar ($R_{23} \text{ and } R_{31}$), by~the symmetry of mapping Equation~\eqref{Eq:qutrit-logical} with respect to cyclic permutations of the three qutrits~\cite{Cleve-1999a}. Realizing the recovery unitary transformation, $R_{ij}$, for~our circuit amounts to numerous SWAP gates (equivalently, three CNOT gates), which results in a much more complex circuit compared to the previous QSS schemes. Furthermore, it should be noted that the construction of the qutrit scheme does not follow the same stabilizer approach as the ((3,5)) and ((5,7)) schemes, and~does not utilize mid-circuit measurements or classical feed-forwarding. The circuit for our qubit version of the qutrit scheme is shown in Figure~\ref{Fig:3-qutrit}.

\section{Results}

\label{Sect:results}

We now present experimental results for the performance of the above scheme using IBM's 127-qubit Brisbane (Eagle r3) system as the experimental platform. We compare our real hardware results with theoretical expectations from simulating noisy circuits using Qiskit's fake backend \textit{FakeBrisbane}, which runs on a classical computer. We also provide noisy simulation results using \textit{FakeTorino}, a~133-qubit (Heron r1) device that uses a different native gate set; simulated results are denoted ``(SIM)''. Unfortunately, we were unable to obtain real results from the newer IBM Torino due to the unavailability of dynamic circuits at the time of data collection. Fake backends are constructed to mimic the behaviors of real IBM quantum devices using system snapshots. The~system snapshots contain important information about the quantum system, such as basis gates, coupling map, gate error rates, and~decoherence times (e.g., $T_1$ and $T_2$), but~do not model dynamic cross-talk or scheduler-induced idling on the day of execution, which limits their fidelity as predictors for deep, feed-forward circuits. The results were taken after the application of readout error mitigation using M3. M3 works with quasi-probability distributions; however, in~our case, a~true probability distribution is a bit more useful. Therefore, we take the returned quasi-probability distribution and map it to the closest probability distribution as defined by the $L_2$-norm.

\begin{figure}[!]
    \begin{subfigure}{0.45\textwidth}
        \caption{}
        \includegraphics[scale=0.3]{SWAP_BRIS_MCM.png}
        \label{Fig:swap_result} 
    \end{subfigure}
    \hfill
    \begin{subfigure}{0.45\textwidth}
        \caption{}
        \includegraphics[scale=0.3]{ENT_BRIS_MCM.png}
        \label{Fig:EF_result}
    \end{subfigure}
\caption[Two numerical solutions]{(a) Using mid-circuit measurements: Sample rate of passing the SWAP test for QSS schemes on the IBM Quantum Brisbane system and corresponding noisy simulators at every optimization level; the size of the erased set is two for the qubit schemes. (b) Entanglement fidelity for QSS schemes using the IBM Quantum Brisbane system and corresponding noisy simulators at every optimization level. For both figures of merit, the ideal value is the noiseless case, meaning the value measured if the schemes are run without noise present. The error bars shown correspond to a $99\%$ confidence interval.}
\label{Fig:result_MCM}
\end{figure}

\begin{figure}[!]
    \begin{subfigure}{0.45\textwidth}
        \caption{}
        \includegraphics[scale=0.3]{SWAP_BRIS_DCM.png}
        \label{Fig:SWAP_BRIS_DCM} 
    \end{subfigure}
    \hfill
    \begin{subfigure}{0.45\textwidth}
        \caption{}
        \includegraphics[scale=0.3]{ENT_BRIS_DCM.png}
        \label{Fig:ENT_BRIS_DCM}
    \end{subfigure}
\caption[Two numerical solutions]{(a) Using delayed circuit measurement: Sample rate of passing the SWAP test for QSS schemes on the IBM Quantum Brisbane system and corresponding noisy simulators at every optimization level; the size of the erased set is two for the qubit schemes. (b) Entanglement fidelity for QSS schemes using the IBM Quantum Brisbane system and corresponding noisy simulators at every optimization level.}
\label{Fig:result_DCM}
\end{figure}

\begin{figure}[!]
    \begin{subfigure}{0.45\textwidth}
        \caption{}
        \includegraphics[scale=0.3]{SWAP_BRIS_MCM_3ER.png}
        \label{Fig:SWAP_BRIS_MCM_3ER} 
    \end{subfigure}
    \hfill
    \begin{subfigure}{0.45\textwidth}
        \caption{}
        \includegraphics[scale=0.3]{ENT_BRIS_MCM_3ER.png}
        \label{Fig:ENT_BRIS_MCM_3ER}
    \end{subfigure}
\caption[Two numerical solutions]{(a) Sample rate of passing the SWAP test for erased sets of size two and three (both $\notin \Gamma$) for the Steane Code at every optimization level. (b) Entanglement fidelity for unauthorized ($\notin \Gamma$) and authorized ($\in \Gamma$) erased sets of size three for the Steane Code at every optimization level. The ideal (noiseless) case is also included (far right) when an authorized set of three qubits is erased. Data is collected using FakeBrisbane and FakeTorino.
}
\end{figure}

\begin{figure}[!]
    \begin{subfigure}{0.45\textwidth}
        \caption{}
        \includegraphics[scale=0.3]{SWAP_TOR.png}
        \label{Fig:tor_swap_result} 
    \end{subfigure}
    \hfill
    \begin{subfigure}{0.45\textwidth}
        \caption{}
        \includegraphics[scale=0.3]{ENT_TOR.png}
        \label{Fig:tor_EF_result}
    \end{subfigure}
\caption[Two numerical solutions]{(a) Sample rate of passing the SWAP test for QSS schemes using FakeTorino with MCM and DCM at every optimization level; the size of the erased set is two for the qubit schemes. (b) Entanglement fidelity for QSS schemes using FakeTorino with MCM and DCM at every optimization level.}
\label{Fig:tor_result}
\end{figure}

For the SWAP test results, five jobs were run for each scheme, each using 20,000 shots. Each job also used arbitrary initialization angles for the quantum states.
We first evaluated the secret recovery by performing a SWAP test as described in Section~\ref{Sect:Performanc_Measures}. The~figure of merit is the sample rate over 20,000 shots of measuring the $\ket{0}$ state in the SWAP test, which occurs with probability one when the initial and final states are identical. We then estimated the entanglement fidelity of the encoding and decoding process for a given subset of erasures using QST. For~QST, five jobs were run for each scheme, but~using \mbox{10,000 shots} each, as~this method is computationally expensive. The results were obtained at each of Qiskit's optimization levels, with 0 meaning no optimization, and~1, 2, and~3 meaning light, medium, and~heavy optimization, respectively.

For the ((3,5)) and ((5,7)) schemes, we evaluated two variations in the circuit: a mid-circuit measurement (MCM) and a delayed-circuit measurement (DCM). MCMs utilize measurements that execute before the end of the circuit, while DCMs only utilize measurements at the end of the circuit.  Finally, for~the Steane code QSS scheme, we evaluated the performance of the code with two and three-qubit erasures. For~the three-qubit erasures, we looked at erasing sets of unauthorized and authorized sets, i.e.,~erasures that do and do not allow~recovery.

We can see from the hardware data in Figure~\ref{Fig:swap_result} that the ((3,5)) and ((5,7)) schemes were the most successful at decoding the secret for the SWAP test, with~the ((5,7)) scheme coming out very slightly ahead among most optimization levels. For~both these schemes, the~simulation predicts much higher results. This large sim--hardware gap (especially in entanglement fidelity) is consistent with three effects already evidenced by our circuits: (i)~idle-time decoherence from circuit depth and mid-circuit synchronization, as~seen for the depth and multi-qubit counts in Appendix \ref{App:Depth}, which disproportionately harms the QST runs that add a reference qubit register; (ii) cross-talk during concurrent two-qubit operations (echoed cross-resonance), which is not fully captured by snapshot-based fake backends; and (iii) residual measurement error and feed-forward scheduling overhead despite M3, especially in the MCM variants. Together, these could explain why the SWAP test pass rates remain reasonable while the reconstructed Bell-pair fidelity collapses more sharply on hardware than in simulation. The~((2,3)) scheme requires qutrit modular-add operations (e.g., the~$R_{ij}$ recovery) that, under~the qubit embedding of Equation~(\ref{Eq:qutrit-encoding}), decompose into chains of SWAP/CNOT (ECR) operations with additional single-qubit rotations and routing (Figure~\ref{Fig:3-qutrit}). After~transpilation, this yields substantially larger two-qubit counts and depth than the qubit codes (see Appendix \ref{App:Depth}, Figure~\ref{Fig:3DepthGates} vs. Figures~\ref{Fig:5DepthGates} and \ref{Fig:7DepthGates}), explaining the systematically worse hardware~performance.

For the qubit schemes, the~MCM construction clearly outperformed the DCM construction by a noticeable margin in both the SWAP test and entanglement fidelity for both real and simulated data, as~shown in Figures~\ref{Fig:result_MCM} and \ref{Fig:result_DCM}. This strongly implies that using classical feed-forward decoding is more efficient than fully coherent decoding with a delayed measurement with these schemes, and~this is supported by the resulting circuit depth and number of multi-qubit operations. Furthermore, as~mentioned previously, we see a large discrepancy between the real and simulated data for the entanglement fidelity of the MCM construction; however, the~discrepancy is not present for the DCM construction, as~both produce the theoretical minimum entanglement fidelity due to the amount of noise/error.

Interestingly, the~Steane code with three erasures slightly outperformed the Steane code with two erasures among most optimization levels concerning the SWAP test, as seen in Figure~\ref{Fig:SWAP_BRIS_MCM_3ER}. This is contrary to what the simulation predicts. Looking at the simulated performance between authorized and unauthorized erased sets of size three for the Steane code, we can see that the erasure of an authorized set drastically reduces the entanglement fidelity. Moreover, it results in the theoretical minimum, as seen in Figure~\ref{Fig:ENT_BRIS_MCM_3ER}. The~results of the ideal or noiseless case further support this~observation.

\section{Conclusions And Outlook}

\label{Sect:conclusion}

In this paper, we provided a detailed procedure for the
construction of encoding and decoding circuits for ((3,5)), ((5,7)), and~((2,3)) quantum secret sharing schemes.
We also went beyond the threshold scheme with the Steane code and investigated its more general access structure, as~well as comparing different measurement implementations (MCM vs. DCM). The~ultimate goal of this work was twofold. First, we wanted to provide a quantum circuit description of quantum secret sharing in an introductory manner that might be useful to newcomers to the field. Second, we wanted to see how well certain QSS schemes perform on a commercial quantum computing platform, such as IBM's 127-qubit Brisbane system. The~scope of this work was just to obtain a benchmarking performance of the IBM machine for the task of secret sharing, as~it is currently available to the~user.

There are multiple reasons for the discrepancies between the ideal, simulated, and~real data. As~noted above, we believe the primary source for non-ideal performance in the SWAP test and entanglement fidelity estimation is the depth of our circuits and, consequently, the~decoherence of idle qubits. A~natural next step would be to employ more tailored error mitigation techniques, such as dynamical decoupling (DD), as~pursued in Ref.~\cite{baumer2024quantumfouriertransformusing}, especially to help with the additional wait time required to perform mid-circuit measurements. Another direction would be to look at machines with different basis (native) gates. Different basis gate sets could improve the performance of the schemes by decreasing the transpilation requirements, as~seen in Figure~\ref{Fig:tor_result}. For~IBM machines specifically, this would require running the schemes on machines with processors other than those with ``Eagle''~processors.

Our results suggest that, for~near-term (pre-fault-tolerance) QSS implementations and for modular settings where secrets/quantum information move between zones, feed-forward-based decoders (MCM) should be preferred to minimize idle exposure. Mid-circuit measurement and feed-forward (MCM) consistently outperform delayed fully coherent variants (DCM) on hardware, aligning with our depth/multi-qubit count comparisons; this favors architectures with fast, low-latency measurement, reset, and~classical control so that syndrome extraction and conditional recovery can proceed with minimal idling. Moreover, platforms that offer native qudit operations (or lower-overhead embeddings/gate sets) or support for higher-dimensional qudits (vs. emulation) would substantially reduce depth and improve robustness for higher-dimensional encodings. Lastly, the~simulation--hardware gap we see is dominated by scheduler-induced idling, cross-talk, and~readout effects that are not fully captured by snapshot-based fake backends; this points to scheduler-aware compilation and scheduler-aware error mitigation (e.g., DD targeted at idle windows) as high-value methods that are likely to yield the biggest marginal gain in fidelity before full fault-tolerance is~achieved.

\begin{acknowledgments}
E.C. is very grateful to Alireza Seif for discussing different error mitigation techniques on IBM machines. We would also like to thank Sarah Hagen for their useful discussions. This research was funded by the National Science Foundation Research Experience for Undergraduates (NSF-REU) under Grant No. PHY-2112890.
\end{acknowledgments}

\clearpage
\bibliography{references}

\newpage

\onecolumngrid
\appendix
\newpage

\section{Syndrome Tables and Corrections}
\label{App:Syndrome-tables}

\begin{table*}[htp]
\centering
\begin{tabular}{ r || c |c |c |c|}
 & $\mbb{I}_2$ & $X_2$ & $Y_2$ & $Z_2$\\
 \hline \hline
 \multirow{2}{1.5em}{$\mbb{I}_1$} & syndrome = (0,0,0,0)& syndrome = (1,0,1,0)&syndrome = (1,1,1,0)&syndrome = (0,1,0,0)\\
 & correction $\mbb{I}_5$ & correction $X_5$ & correction $X_5$ & correction $\mbb{I}_5$\\
 \hline
  \multirow{2}{1.5em}{$X_1$} & syndrome = (1,0,1,1)& syndrome = (0,0,0,1)&syndrome = (0,1,0,1)&syndrome = (1,1,1,1)\\
 & correction $Y_5$ & correction $Z_5$ & correction $Z_5$ & correction $Y_5$\\
 \hline
  \multirow{2}{1.5em}{$Y_1$} & syndrome = (0,0,1,1)& syndrome = (1,0,0,1)&syndrome = (1,1,0,1)&syndrome = (0,1,1,1)\\
 & correction $Y_5$ & correction $Z_5$ & correction $Z_5$ & correction $Y_5$\\
 \hline
  \multirow{2}{1.5em}{$Z_1$} & syndrome = (1,0,0,0)& syndrome = (0,0,1,0)&syndrome = (0,1,1,0)&syndrome = (1,1,0,0)\\
& correction $\mbb{I}_5$ & correction $X_5$ & correction $X_5$ & correction $\mbb{I}_5$\\
\end{tabular}
\caption{For arbitrary Pauli errors on qubits one and two in the $((3,5))$ QSS scheme, this table provides the unitary correction $R_k$ on qubit five for each syndrome measurement $\mbf{b}_k$.}
\label{Tab:correction-5-qubit}
\end{table*}
\begin{table*}[htp]
\centering
\begin{tabular}{ r || c |c |c |c|}
 & $\mbb{I}_7$ & $X_7$ & $Y_7$ & $Z_7$\\
 \hline \hline
 \multirow{2}{1.5em}{$\mbb{I}_6$} & syndrome = (0,0,0,0,0,0)& syndrome = (0,1,1,0,0,0)&syndrome = (0,1,1,0,0,1)&syndrome = (0,0,0,0,0,1)\\
 & correction $\mbb{I}_1$ & correction $X_1$ & correction $X_1$ & correction $\mbb{I}_1$\\
 \hline
  \multirow{2}{1.5em}{$X_6$} & syndrome = (1,0,1,0,0,0)& syndrome = (1,1,0,0,0,0)&syndrome = (1,1,0,0,0,1)&syndrome = (1,0,1,0,0,1)\\
 & correction $X_1$ & correction $\mbb{I}_1$ & correction $\mbb{I}_1$ & correction $X_1$\\
 \hline
  \multirow{2}{1.5em}{$Y_6$} & syndrome = (1,0,1,0,1,0)& syndrome = (1,1,0,0,1,0)&syndrome = (1,1,0,0,1,1)&syndrome = (1,0,1,0,1,1)\\
 & correction $X_1$ & correction $\mbb{I}_1$ & correction $\mbb{I}_1$ & correction $X_1$\\
 \hline
  \multirow{2}{1.5em}{$Z_6$} & syndrome = (0,0,0,0,1,0)& syndrome = (0,1,1,0,1,0)&syndrome = (0,1,1,0,1,1)&syndrome = (0,0,0,0,1,1)\\
& correction $\mbb{I}_1$ & correction $X_1$ & correction $X_1$ & correction $\mbb{I}_1$\\
\end{tabular}
\caption{For arbitrary Pauli errors on qubits six and seven in the 7-qubit QSS scheme, this table provides the unitary correction $R_k$ on qubit one for each syndrome measurement $\mbf{b}_k$.}
\label{Tab:correction-7-qubit}
\end{table*}

\newpage
\section{[[7,1,3]] Steane code and three-qubit erasure lookup table}
\label{App:Steane-access}

\begin{figure}[h]
        \includegraphics[scale=0.33]{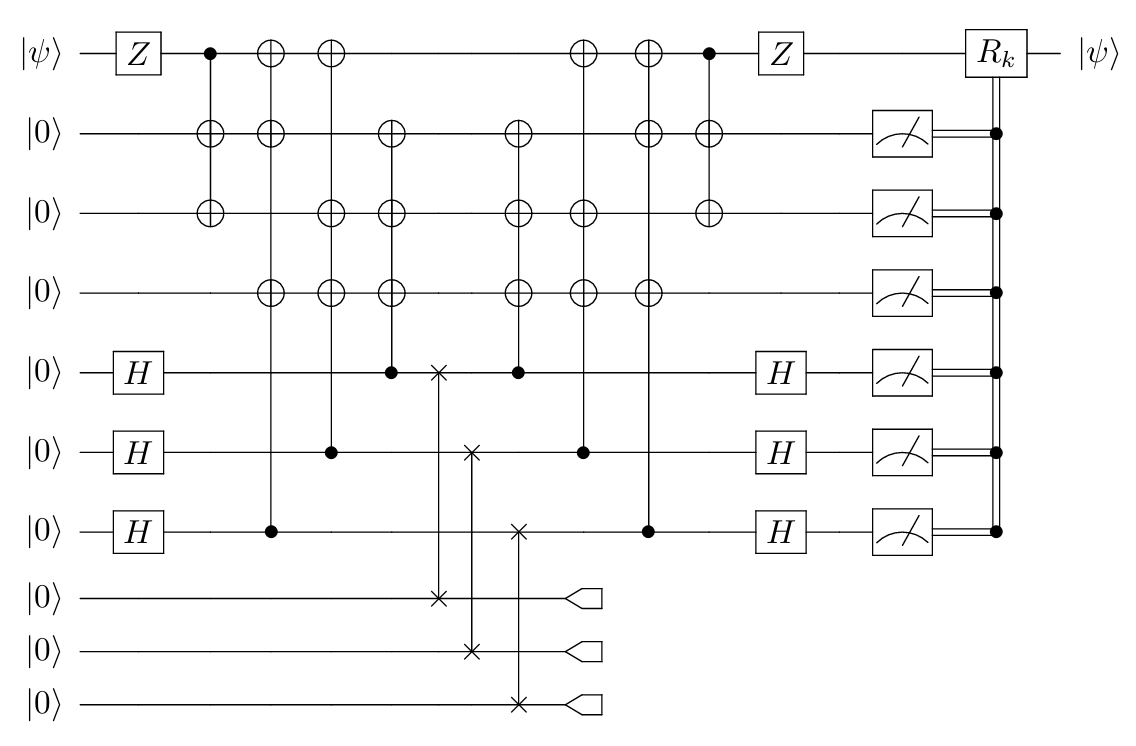}
\caption[Two numerical solutions]{A circuit implementation of a Seven-qubit (Steane) QSS protocol with three qubits being erased, $|E|=3$, $E\notin \Gamma$.}
\label{Fig:4-7-qubit-circuit}
\end{figure}

The Steane code can also recover from $d=3$ erasures of certain qubits. One such recovery is depicted in Figure~\ref{Fig:4-7-qubit-circuit}. We further verify below that a subset of three qubits exists in a maximally mixed state, which implies that these qubits are independent of the logical state. Consequently, the~remaining four qubits must contain all the information about the original encoded quantum~state.

The encoded state is
\begin{equation}
\label{append_1}
\ket{\psi}_L = \frac{\alpha}{\sqrt{8}} \sum_{i=1}^8 \ket{u_i} + \frac{\beta}{\sqrt{8}} \sum_{j=1}^8 \ket{v_j},
\end{equation}
where
\begin{align}
\ket{u_i}\in\{&
\ket{0000000},
\ket{1010101},
\ket{0110011},
\ket{1100110}, \notag\\
&\ket{0001111},
\ket{1011010},
\ket{0111100},
\ket{1101001}\}\notag\\
\ket{v_j}\in\{&
\ket{1111111},
\ket{0101010},
\ket{1001100},
\ket{0011001}, \notag \\
&\ket{1110000},
\ket{0100101},
\ket{1000011},
\ket{0010110}\}.
\end{align}
We 
 denote the encoded density matrix as $\rho:= \op{\psi}{\psi}_L$.

To find the reduced density matrix for the set of qubits, $S=\{q_5,q_6,q_7\}$, we must first trace out $\overline{S} = \{q_1,q_2,q_3,q_4\}$ from the encoded state $\ket{\psi}_L$:
\begin{equation}
\label{append_3}
\rho^{S} = \tr_{\overline{S}} (\op{\psi}{\psi}_L) = \sum_{i,j,k,l=0}^1 \langle ijkl | \rho | ijkl \rangle \\.
\end{equation}
We find that
\begin{equation}
\label{append_4}
\rho^{S} = \left( \frac{|\alpha|^2 + |\beta|^2}{8} \right) \sum_{p,q,r=0}^1 \op{pqr}{pqr} = \frac{1}{8}\mbb{I}_{8},
\end{equation}
where $\mbb{I}_8$ is the $8\times8$ identity matrix. This implies that the other four qubits carry \textit{all} the information about the quantum state, and~thus, four qubits can be used to recover the secret. Table~\ref{table:correction-4-7-qubit} provides the correction for each measured syndrome. On~the other hand, there are certain three-qubit erasures that cannot be corrected. For~example, consider the erasure of qubits $\{q_2,q_4,q_6\}$. A~straightforward calculation shows that the reduced density matrix after tracing out these qubits has some dependence on the encoded state $\ket{\psi}$.

\begin{table*}[h]
\caption{For arbitrary Pauli errors on qubits five, six, and seven in the Steane QSS scheme, this table provides the unitary correction $R_k$ on qubit one for each syndrome measurement $\mbf{b}_k$.} 
\centering 
\begin{ruledtabular}
\begin{tabular}{c c c c c c} 
Error & Syndrome & Correction & Error & Syndrome & Correction \\  
\hline 
$\mathbb{I}_5 \mathbb{I}_6 \mathbb{I}_7$ & $(0,0,0,0,0,0)$ & $\mathbb{I}_1$ & $Y_5 \mathbb{I}_6 \mathbb{I}_7$ & $(1,1,1,1,0,0)$ & $\mathbb{I}_1$ \\ 
$\mathbb{I}_5 \mathbb{I}_6 X_7$ & $(0,1,1,0,0,0)$ & $X_1$ & $Y_5 \mathbb{I}_6 X_7$ & $(1,0,0,1,0,0)$ & $X_1$ \\
$\mathbb{I}_5 \mathbb{I}_6 Y_7$ & $(0,1,1,0,0,1)$ & $X_1$ & $Y_5 \mathbb{I}_6 Y_7$ & $(1,0,0,1,0,1)$ & $X_1$ \\ 
$\mathbb{I}_5 \mathbb{I}_6 Z_7$ & $(0,0,0,0,0,1)$ & $\mathbb{I}_1$ & $Y_5 \mathbb{I}_6 Z_7$ & $(1,1,1,1,0,1)$ & $\mathbb{I}_1$ \\
$\mathbb{I}_5 X_6 \mathbb{I}_7$ & $(1,0,1,0,0,0)$ & $X_1$ & $Y_5 X_6 \mathbb{I}_7$ & $(0,1,0,1,0,0)$ & $X_1$ \\ 
$\mathbb{I}_5 X_6 X_7$ & $(1,1,0,0,0,0)$ & $\mathbb{I}_1$ & $Y_5 X_6 X_7$ & $(0,0,1,1,0,0)$ & $\mathbb{I}_1$ \\ 
$\mathbb{I}_5 X_6 Y_7$ & $(1,1,0,0,0,1)$ & $\mathbb{I}_1$ & $Y_5 X_6 Y_7$ & $(0,0,1,1,0,1)$ & $\mathbb{I}_1$ \\ 
$\mathbb{I}_5 X_6 Z_7$ & $(1,0,1,0,0,1)$ & $X_1$ & $Y_5 X_6 Z_7$ & $(0,1,0,1,0,1)$ & $X_1$ \\ 
$\mathbb{I}_5 Y_6 \mathbb{I}_7$ & $(1,0,1,0,1,0)$ & $X_1$ & $Y_5 Y_6 \mathbb{I}_7$ & $(0,1,0,1,1,0)$ & $X_1$ \\
$\mathbb{I}_5 Y_6 X_7$ & $(1,1,0,0,1,0)$ & $\mathbb{I}_1$ & $Y_5 Y_6 X_7$ & $(0,0,1,1,1,0)$ & $\mathbb{I}_1$ \\
$\mathbb{I}_5 Y_6 Y_7$ & $(1,1,0,0,1,1)$ & $\mathbb{I}_1$ & $Y_5 Y_6 Y_7$ & $(0,0,1,1,1,1)$ & $\mathbb{I}_1$ \\
$\mathbb{I}_5 Y_6 Z_7$ & $(1,0,1,0,1,1)$ & $X_1$ & $Y_5 Y_6 Z_7$ & $(0,1,0,1,1,1)$ & $X_1$ \\
$\mathbb{I}_5 Z_6 \mathbb{I}_7$ & $(0,0,0,0,1,0)$ & $\mathbb{I}_1$ & $Y_5 Z_6 \mathbb{I}_7$ & $(1,1,1,1,1,0)$ & $\mathbb{I}_1$ \\
$\mathbb{I}_5 Z_6 X_7$ & $(0,1,1,0,1,0)$ & $X_1$ & $Y_5 Z_6 X_7$ & $(1,0,0,1,1,0)$ & $X_1$ \\
$\mathbb{I}_5 Z_6 Y_7$ & $(0,1,1,0,1,1)$ & $X_1$ & $Y_5 Z_6 Y_7$ & $(1,0,0,1,1,1)$ & $X_1$ \\
$\mathbb{I}_5 Z_6 Z_7$ & $(0,0,0,0,1,1)$ & $\mathbb{I}_1$ & $Y_5 Z_6 Z_7$ & $(1,1,1,1,1,1)$ & $\mathbb{I}_1$ \\
$X_5 \mathbb{I}_6 \mathbb{I}_7$ & $(1,1,1,0,0,0)$ & $\mathbb{I}_1$ & $Z_5 \mathbb{I}_6 \mathbb{I}_7$ & $(0,0,0,1,0,0)$ & $\mathbb{I}_1$ \\ 
$X_5 \mathbb{I}_6 X_7$ & $(1,0,0,0,0,0)$ & $X_1$ & $Z_5 \mathbb{I}_6 X_7$ & $(0,1,1,1,0,0)$ & $X_1$ \\
$X_5 \mathbb{I}_6 Y_7$ & $(1,0,0,0,0,1)$ & $X_1$ & $Z_5 \mathbb{I}_6 Y_7$ & $(0,1,1,1,0,1)$ & $X_1$ \\
$X_5 \mathbb{I}_6 Z_7$ & $(1,1,1,0,0,1)$ & $\mathbb{I}_1$ & $Z_5 \mathbb{I}_6 Z_7$ & $(0,0,0,1,0,1)$ & $\mathbb{I}_1$ \\
$X_5 X_6 \mathbb{I}_7$ & $(0,1,0,0,0,0)$ & $X_1$ & $Z_5 X_6 \mathbb{I}_7$ & $(1,0,1,1,0,0)$ & $X_1$ \\
$X_5 X_6 X_7$ & $(0,0,1,0,0,0)$ & $\mathbb{I}_1$ & $Z_5 X_6 X_7$ & $(1,1,0,1,0,0)$ & $\mathbb{I}_1$ \\
$X_5 X_6 Y_7$ & $(0,0,1,0,0,1)$ & $\mathbb{I}_1$ & $Z_5 X_6 Y_7$ & $(1,1,0,1,0,1)$ & $\mathbb{I}_1$ \\
$X_5 X_6 Z_7$ & $(0,1,0,0,0,1)$ & $X_1$ & $Z_5 X_6 Z_7$ & $(1,0,1,1,0,1)$ & $X_1$ \\
$X_5 Y_6 \mathbb{I}_7$ & $(0,1,0,0,1,0)$ & $X_1$ & $Z_5 Y_6 \mathbb{I}_7$ & $(1,0,1,1,1,0)$ & $X_1$ \\
$X_5 Y_6 X_7$ & $(0,0,1,0,1,0)$ & $\mathbb{I}_1$ & $Z_5 Y_6 X_7$ & $(1,1,0,1,1,0)$ & $\mathbb{I}_1$ \\ 
$X_5 Y_6 Y_7$ & $(0,0,1,0,1,1)$ & $\mathbb{I}_1$ & $Z_5 Y_6 Y_7$ & $(1,1,0,1,1,1)$ & $\mathbb{I}_1$ \\
$X_5 Y_6 Z_7$ & $(0,1,0,0,1,1)$ & $X_1$ & $Z_5 Y_6 Z_7$ & $(1,0,1,1,1,1)$ & $X_1$ \\
$X_5 Z_6 \mathbb{I}_7$ & $(1,1,1,0,1,0)$ & $\mathbb{I}_1$ & $Z_5 Z_6 \mathbb{I}_7$ & $(0,0,0,1,1,0)$ & $\mathbb{I}_1$ \\
$X_5 Z_6 X_7$ & $(1,0,0,0,1,0)$ & $X_1$ & $Z_5 Z_6 X_7$ & $(0,1,1,1,1,0)$ & $X_1$ \\
$X_5 Z_6 Y_7$ & $(1,0,0,0,1,1)$ & $X_1$ & $Z_5 Z_6 Y_7$ & $(0,1,1,1,1,1)$ & $X_1$ \\
$X_5 Z_6 Z_7$ & $(1,1,1,0,1,1)$ & $\mathbb{I}_1$ & $Z_5 Z_6 Z_7$ & $(0,0,0,1,1,1)$ & $\mathbb{I}_1$ \\ 
\end{tabular}
\end{ruledtabular}
\label{table:correction-4-7-qubit} 
\end{table*}

\clearpage
\section{Depth \& Multi-qubit operations for the QSS Schemes}
\label{App:Depth}

\begin{figure*}[htp]
\centering
\captionsetup{justification=centering}
  \includegraphics[scale=0.5]{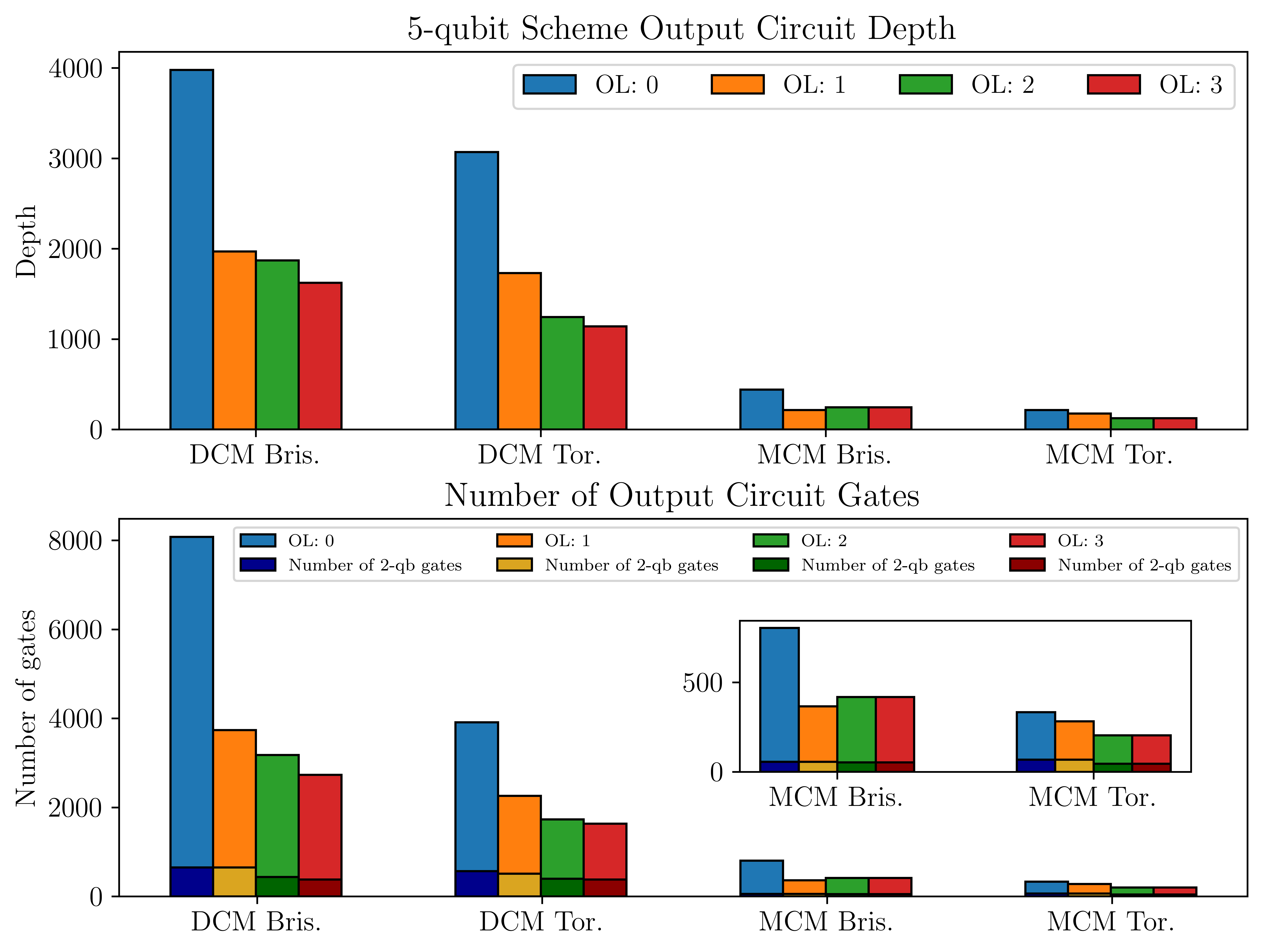}
  \caption{Transpiled circuit depth and gate count for ((3,5)) scheme}
  \label{Fig:5DepthGates}
\end{figure*}

\begin{figure*}[htp]
\centering
\captionsetup{justification=centering}
  \includegraphics[scale=0.5]{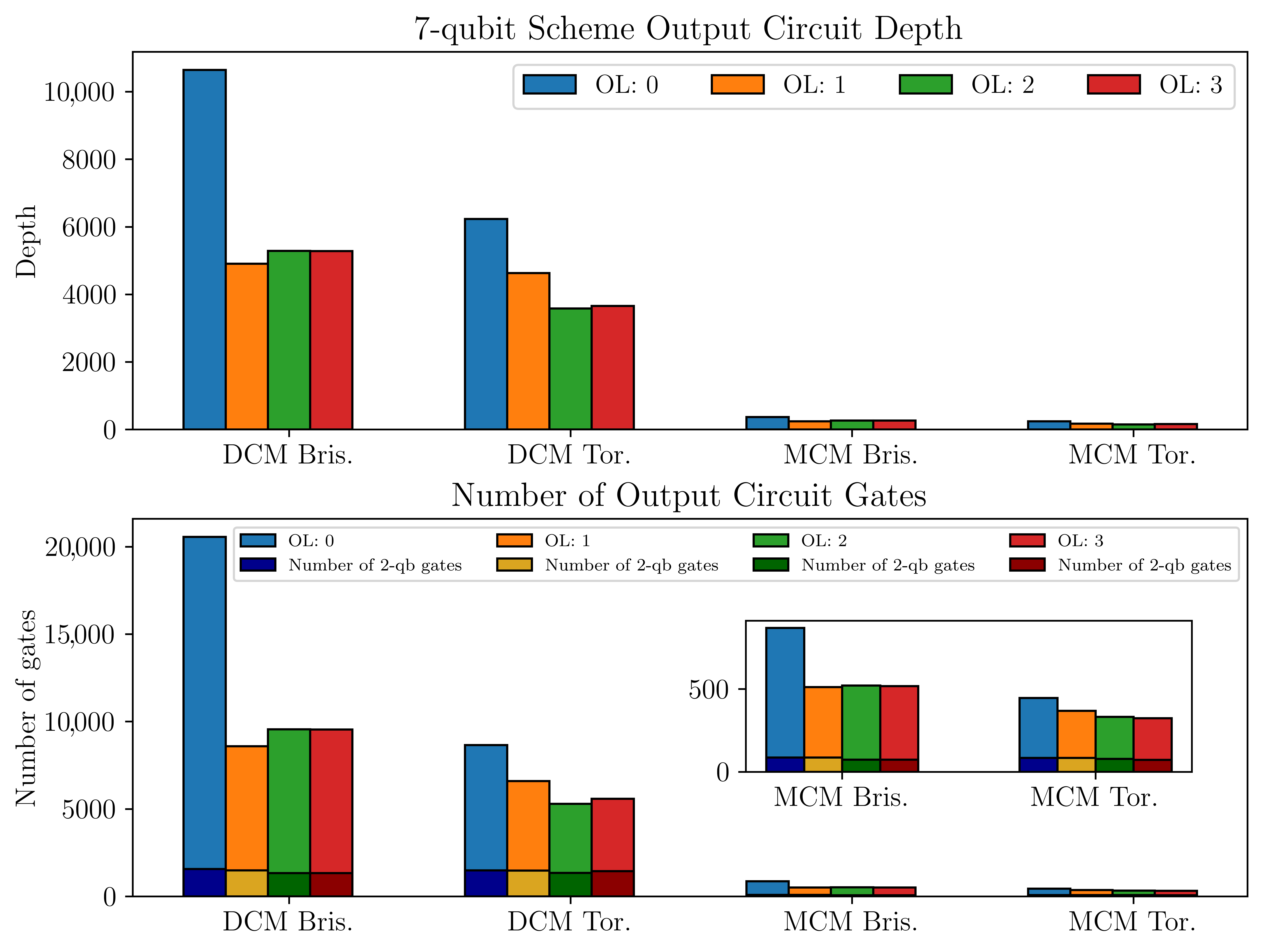}
  \caption{Transpiled circuit depth and gate count for ((5,7)) scheme}
  \label{Fig:7DepthGates}
\end{figure*}

\begin{figure*}[htp]
\centering
\captionsetup{justification=centering}
  \includegraphics[scale=0.5]{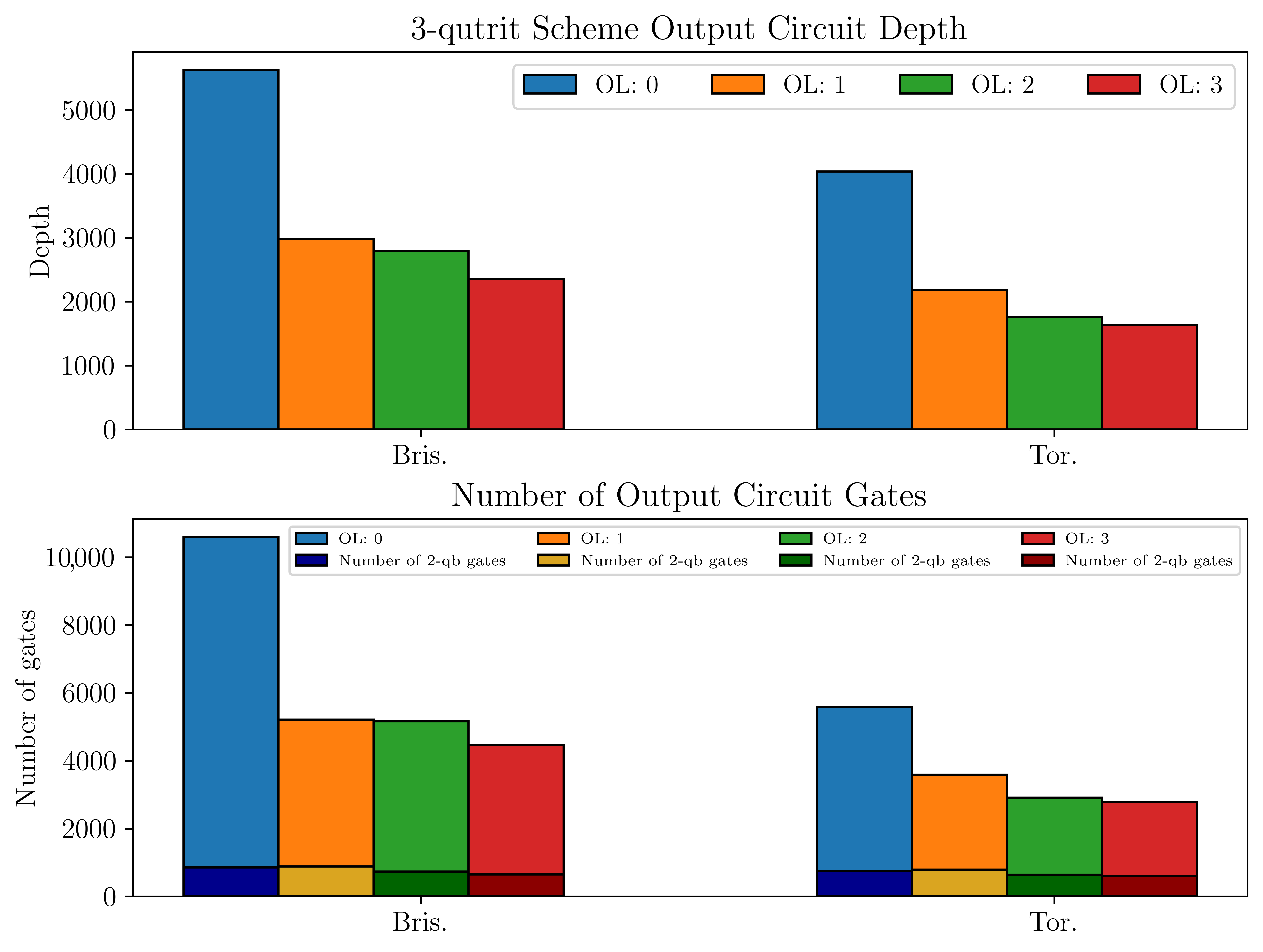}
  \caption{Transpiled circuit depth and gate count for ((2,3)) scheme}
  \label{Fig:3DepthGates}
\end{figure*}

\end{document}